# A Combined Dependability and Security Approach for Third Party Software in Space Systems

David Escorial Rico
Product Assurance and Safety Department
ESA-ESTEC
Noordwijk, The Netherlands

Mark Hann
Space Division
SCISYS UK Limited
Bristol, United Kingdom

*Abstract*— **Software components for on-board architectures in the space domain are increasingly reliant on Commercial Off-The-Shelf (COTS), Open Source (OSS) or other third party software products. However, these software components often have not been built with mission critical requirements in mind. Development project teams incorporating these products have limited knowledge of or control over the processes applied during the design, implementation, testing and maintenance of selected COTS/OSS software products. These constraints generate uncertainty of potential software induced failures. Moreover, the lack of information regarding security vulnerabilities increases the risks of their usage, since their exploitation might lead to undesired behaviour of the software and therefore to a system failure. The purpose of this paper is to present a combined approach that takes into account reliability and security enhancements for third party software, based on Time-Space Partitioning and Multiple Levels of Security.**

## I. INTRODUCTION

Space systems have been increasingly relying on software components for all kind of tasks: commanding, telemetry collection, attitude and orbit control, payload management, thermal control, etc. A complex spacecraft such as the Automated Transfer Vehicle (ATV) consists of up to one million lines of code, the flight application software contains 450 Kloc [1]. Because of the large size of software systems, it is not uncommon that they include not only reused components of previous missions, but third party components developed outside supervision of the project team, for example, Commercial Off The Shelf (COTS) or even Open Source Software (OSS). One of the main challenges of the development of space systems with third party software products is to provide confidence that those software components perform as specified in their requirements within the target operational environment [2]. Considering the uncertainty caused by the lack of development visibility of these software components and the potential for failure, this essentially implies meeting Dependability requirements. For this purpose, there exists a variety of methods for fault prevention, tolerance, removal and forecasting. However, software Security is concerned with integrity, confidentiality and availability requirements. Due to the inability to assure the compliance of externally developed software components with these requirements, an additional set of Security methods, such as partitioning, ciphering and vulnerability prevention, could be included in the system design.

The important aspect to take into account is that both Security and Dependability are strongly interdependent, generating opportunities of complementarity and potential trade-offs. There are approaches that can be used during the development of the software in order to assure required levels of Security and Dependability when third party software is present. In this paper we will discuss containers, safety/security boxes, inhibitors, monitors and how they can be combined.

## II. A COMBINED APPROACH FOR SECURITY AND DEPENDABILITY: THE SEPARATION KERNEL

A previous study about Software Failure Propagation Prevention [3] analysed methods to prevent failure propagation between on board software components and to other parts of the system. It concluded that the most appropriate technology to address failure propagation, namely due to resource hogging and resource violation failures, is the use of a separation kernel. Separation kernels provide Time and Space Partitioning (TSP) [4][5] of applications into their own partitions which prevents failure propagation via the unplanned use of shared resources of CPU and memory [6]. They also implement the concept of Multiple Independent Levels of Security (MILS) that was developed to simplify monolithic security kernels enforcing security policies [7], allow controlled communications between non-kernel partitions, and enable the applications to enforce their own security policies.

Separation kernels provide a safety/security box for third party software products. All the software in the partition containing the third party software is designated as having the lowest Dependability and Security assurance level. As all the software in a partition shares the same resources, any failures can be propagated between software components in that specific partition but not to other partitions containing higher criticality software functions. Therefore, the separation kernel provides the core component of the trusted computing base which enforces the Dependability and Security properties of the system.

In the context of the presence of third party (COTS/OSS) software, the separation kernel needs to be configured taking into account these specific requirements:

- Memory of the other partitions must not be visible or writable by the COTS/OSS software partition. There must not be any shared memory resources between partitions.
- Information flow from one partition to another must be from an authenticated source to authenticated recipients; the source of information is authenticated to the recipient, and information goes only where intended.
- It must be ensured that the microprocessor and any networking equipment cannot be used as a covert channel to leak information.
- Damage is limited by preventing a failure in one partition from cascading to any other partition. Failures are detected, contained, and recovered locally.

In addition to the separation kernel the design must include a number of elements to support the requirements: A Communication Controller, Confidentiality and Dependability guards and Monitors.

The Communication Controller manages the routing of the communication between the COTS/OSS partition and other on board applications or externally. It is functionally similar to a router for inter-partition communication within a Time and Space Partitioned system. It protects against unauthorized transmission of data between applications of different levels of confidentiality. However, it also must allow reconfiguration of information flows by modifying the data tables in one partition.

Guards prevent confidential information, unreliable information or empty information being passed to another application.

Monitors observe independently the operation of the COTS/OSS in its partition. They are either active, where they cyclically check the operation of the OSS, or passive, where they respond to a failure event in the COTS/OSS partition. When a failure is detected they start appropriate recovery actions.

The Communication Controller, the Monitors and the Guards must be independent, so they must not be located in the COTS/OSS partition, but in the so called System partition.

## III. CONCLUSIONS

The concept of combined dependability and security techniques based on a separation kernel have been demonstrated considering third party software products [8].

It should be noted that there was no need to perform a trade-off between Security and Dependability requirements except when defining the reliability and confidentiality guards. Some checks, such as data structure and consistency, might depend on application knowledge. If they were placed in the application that produced the data itself then they would be performed after the data is passed through the Communication Controller and confidentiality guards. As this could potentially cause a conflict with the confidentiality guards, it is recommended that the reliability and availability checks are performed by the application that receives the data.

The components of the System Partition, in particular the confidentiality and dependability Guards, are of static nature, depending on the applications run in the different partitions, therefore if the different applications are uploaded, they should be modified (via patching of the System Partition itself).

A demonstration for on-board image processing using a COTS library was developed in order to prove the concept. This demonstration identified a number of performance effects. First, due to the use of Inter-Partition Communication, copying extra data is required (10.5 % CPU time measured in the demonstrator), and there is additional idle time between partition activations. Moreover the execution of a central Communication Controller implies that further Inter-Partition Communication is needed (4.2 % CPU time). In the case of software maintenance, third party elements need to be uploaded in full, this requires a significantly longer time to restart a partition than a usual maintenance operation, where partial patching can be performed. Finally, there is additional memory required for the separation kernel and the guest operating systems needed to run the different third party applications.

The decision whether to use any third party software component or develop it completely new should also be guided by the impact in the complexity of the overall design. Although not part of the objectives of the current research, it seems reasonable that COTS/OSS products with limited assurance knowledge might be most suitable for providing on board processing to supply higher order data products that are not critical to the spacecraft's operation.

REFERENCES

[1] ATV-3 Edoardo Amaldi, Information Kit, ESA ERASMUS Centre - Directorate of Human Spaceflight and Operation, ESA-HSO-COU-023, rev. 2.0, 2012.

[2] European Cooperation for Space Standardisation. ECSS-Q-ST-80C – Space Product Assurance – Software Product Assurance, March 2009.

[3] M. Hann, P. Rodriguez Dapena, D. Moretti. "Software Failure Propagation Prevention: Application of the Guidelines and Recommendations of Proof of Concept Pilots for Ground and Flight Software", DASIA 2014 - DAta Systems In Aerospace, Warsaw, 2014.

[4] ARINC. Avionics Application Software Standard Interface — ARINC Specification 653-1, October 2003.

[5] A. Esquinas, J. Zamorano, J. A. de la Puente, M. Masmano, A. Crespo. "Time and Space Partition Platform for Safe and Secure Flight Software". Data Systems in Aerospace — DASIA 2012 - DAta Systems In Aerospace, Dubrovnik, 2012.

[6] T. Pareaud. "Securely Partitioned Spacecraft Computing Resources", Final Report, SecPar: Securely Partitioning Spacecraft Computing Resources ESA study, SecPar-FR, August 2011.

[7] J. Alves-foss, P. Oman, W. Scott Harrison, C. Taylor. "The MILS architecture for high-assurance embedded systems". International Journal of Embedded Systems, Vol. 2, Iss. 3/4, 2006

[8] M. Hann. "Demonstrator of a combined Dependability and Security approach for COTS software in Space Systems", Final Report, Demonstrator of a combined Dependability and Security approach for COTS software in Space Systems ESA Study, SSL/10152/DOC/011, Issue 1.1, February 2015.